\documentclass[journal=aamick,manuscript=letter,layout=10pt]{achemso}

\usepackage[version=3]{mhchem} 
\usepackage[T1]{fontenc}       
\usepackage{mciteplus}
\usepackage{hyperref}
\usepackage{amssymb}
\usepackage[all]{hypcap}
\usepackage{natbib}
\usepackage[scaled]{helvet}

\author{Hasan Atesci}
\affiliation{Huygens-Kamerlingh Onnes Laboratorium, Leiden University, P.O. Box  9504, 2300 RA Leiden, The Netherlands}
\author{Francesco Coneri}
\affiliation{MESA+ Institute for Nanotechnology, University of Twente, P.O. Box 217, 7500 AE Enschede, The Netherlands}
\author{Maarten Leeuwenhoek}
\affiliation{Huygens-Kamerlingh Onnes Laboratorium, Leiden University, P.O. Box 9504, 2300 RA Leiden, The Netherlands}
\author{Jouri Bommer}
\affiliation{MESA+ Institute for Nanotechnology, University of Twente, P.O. Box 217, 7500 AE Enschede, The Netherlands}
\author{James R. T. Seddon}
\affiliation{MESA+ Institute for Nanotechnology, University of Twente, P.O. Box 217, 7500 AE Enschede, The Netherlands}
\author{Hans Hilgenkamp}
\affiliation{MESA+ Institute for Nanotechnology, University of Twente, P.O. Box 217, 7500 AE Enschede, The Netherlands}
\author{Jan M. Van Ruitenbeek}
\affiliation{Huygens-Kamerlingh Onnes Laboratorium, Leiden University, P.O. Box 9504, 2300 RA Leiden, The Netherlands}
\email{ruitenbeek@physics.leidenuniv.nl}

\title{On the formation of a conducting surface channel by ionic liquid gating of an insulator}


\begin{document}

\begin{abstract}

Ionic liquid gating has become a popular tool for tuning the charge carrier densities of complex oxides. Among these, the band insulator SrTiO$_3$ is one of the most extensively studied materials. While experiments have succeeded in inducing (super)conductivity, the process by which ionic liquid gating turns this insulator into a conductor is still under scrutiny. Recent experiments have suggested an electrochemical rather than electrostatic origin of the induced charge carriers. Here, we report experiments probing the time evolution of conduction of SrTiO$_3$ near the glass transition temperature of the ionic liquid. By cooling down to temperatures near the glass transition of the ionic liquid the process develops slowly and can be seen to evolve in time. The experiments reveal a process characterized by waiting times that can be as long as several minutes preceding a sudden appearance of conduction. For the conditions applied in our experiments we find a consistent interpretation in terms of an electrostatic mechanism for the formation of a conducting path at the surface of SrTiO$_3$. The mechanism by which the conducting surface channel develops relies on a nearly homogeneous lowering of the surface potential until the conduction band edge of SrTiO$_3$ reaches the Fermi level of the electrodes.

\end{abstract}

\section{Introduction}

The carrier density in materials is the central factor in all electron transport properties. Controlling this carrier density by externally applied gate potentials permits the study of transport and electron-electron interaction effects as a continuous function of this density. Applying an electrostatic potential by means of metallic gates separated by a dielectric from the material under study  allows covering only a limited range of carrier densities. Using ionic liquids (ILs) or ionic gels as dynamic dielectrics between the gate and the device gives access to a much wider range of carrier densities \cite{Ye2009,Shimotani2007,Goldman2014}. When a potential difference is applied across the IL, the entire potential drop is concentrated at the  IL/electrode surfaces, where electric double-layers (EDLs) are formed between a sheet of ions of one dominant polarity and a sheet of induced charges just below the surface of the material. These charge layers are separated by a distance as small as $1$~nm, thereby producing extremely large electric fields. This makes it possible to induce carrier densities as high as $8\cdot 10^{14}$~cm$^{-2}$ at the liquid/solid interface for a material that is otherwise a deep insulator \cite{Ye2009}. These carrier densities are large enough to permit controlling fundamental phenomena such as magnetic order, phase transitions and superconductivity  by means of a gate potential. \cite{Shimotani2007,Ueno2011,Leng2011,Ye2010,Lee2011,Nakano2012}

Among the materials used for such studies strontium titanate, SrTiO$_3$, stands out as one of the most widely studied and best characterized systems \cite{Pai2017}. It is a band insulator with an indirect bandgap of $3.25$~eV, which can be converted into a good conductor at its surface, and even into a superconductor \cite{Ueno2008}. Disorder plays a role, and at low densities the conductance is best described by variable range hopping and the formation of percolation paths for the electrons in the disorder potential \cite{Li2012}. The origin of this disorder could be intrinsic to SrTiO$_3$ \cite{Pai2017}, or it may result from density fluctuations in the ionic liquid \cite{Gallagher2015,Petach2017}. The role of the ionic liquid also comes into play when discussing the nature of gating in terms of a purely electrostatic effect, or (partially) as a result of electrochemical modifications of the surface. For sufficiently strong electric fields and field gradients at the interface one should anticipate inducing disorder and chemical modifications at the surface. Indeed, experiments have shown the influence of IL gating on the oxygen content of complex oxides \cite{Jeong2013,Li2013,Ueno2010}, while others have shown an unusually high buildup of charges \cite{Garcia2013}, characteristic of induced electrochemical reactions. For the material of interest here, SrTiO$_3$, several reports have given conflicting views on the mechanism of IL gating being either electrochemical \cite{Li2013} or electrostatic \cite{Ueno2008,Goldman2014,Gallagher2015} in origin. 

More recently, the dynamics of the formation of conducting surface channels has received attention from several groups \cite{Tsuchiya2015,Schmidt2016,Li2017}. The standard descriptions borrowed from electrochemistry, in terms of homogeneous Helmholtz or Gouy-Chapman layers at the interface with the solid, may not be applicable \cite{Schmidt2016}. A concrete model of the dynamic evolution of a conducting channel was proposed by Tsuchiya {\it et al.} \cite{Tsuchiya2015}, in terms of a gradual spreading of the conducting channel over the surface of the solid, starting from the source and drain contacts. Here we address this question of how the electric double layer forms on an insulator. We therefore seek to investigate the development and underlying mechanism of conductivity across the surface of SrTiO$_3$ in the time domain. To this end, we slow down the gating process by lowering the temperature at which the gate potential is applied, to temperatures just above the glass  transition of the ionic liquid. This allows us to follow the process of gating in time and elucidate the role of the ionic liquid in the process. We were further motivated to investigate the charging dynamics at low temperatures because under such conditions unwanted electrochemical processes are more easily suppressed. Our observations show that, at variance with the interpretation by Tsuchiya {\it et al.}, \cite{Tsuchiya2015} the surface conducting channel on SrTiO$_3$ develops homogeneously. The observed time delay of conductance after switching on the gate potential is attributed to the time required for bringing the surface potential to the conduction band edge of SrTiO$_3$. The subsequent evolution of the conductance of the channel is consistent with a percolation description of the two-dimensional electron system.

\section{Experimental techniques}
\begin{figure}[t!]
	\centering
	\includegraphics[width=0.9\textwidth]{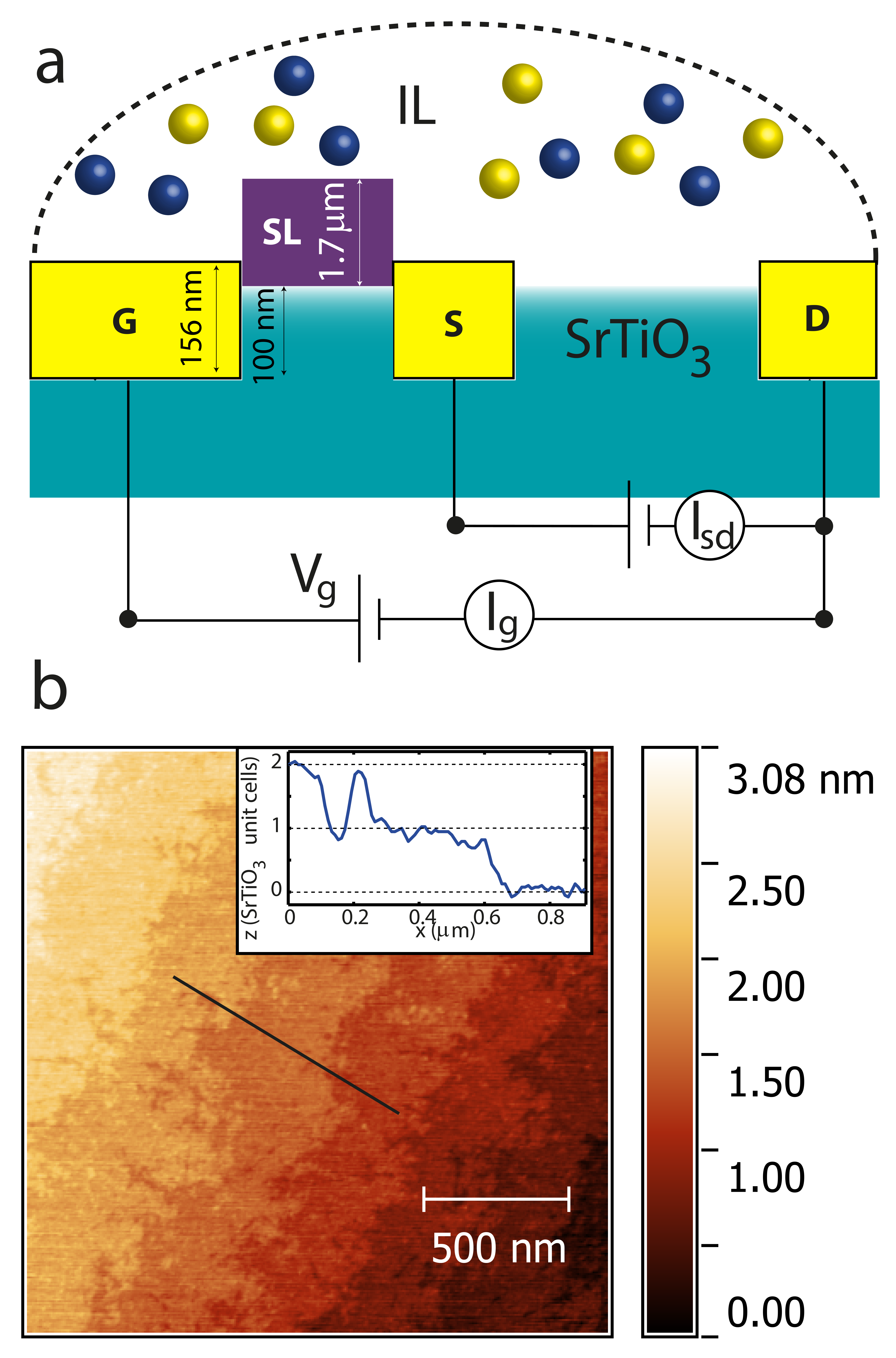}
	\caption{(a) A schematic cross section of the sample. Yellow and blue spheres represent TFSI(-) and DEME(+) ions, respectively. The gate (G), source (S) and drain (D) electrodes have a thickness of $6$ nm Ti plus $150$ nm Au. For the areas onto which the Ti/Au layer is deposited, the SrTiO$_3$ is ion milled to about $100$ nm depth. The separator layer (SL) is $1.7$~$\mu$m thick. (b) AFM image of a typical SrTiO$_3$ surface after an oxygen plasma treatment. Clear terrace steps can be observed. The inset shows the profile indicated by the black line, showing step heights corresponding to the SrTiO$_3$ unit cell height.}
\label{fig:Micrograph}
\end{figure}

\begin{figure}[t!]
	\centering
	\includegraphics[width=0.95\textwidth]{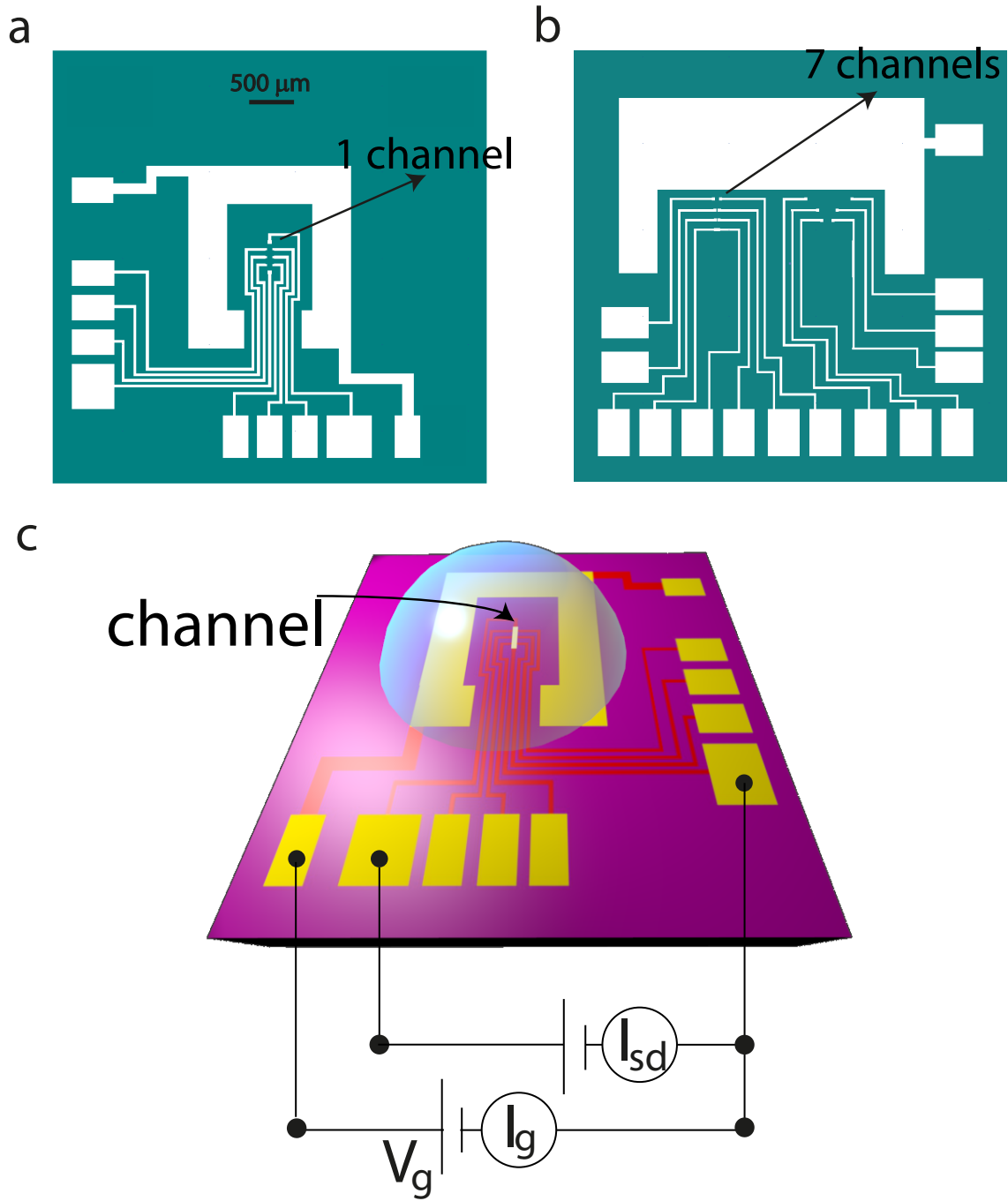}
	\caption{(a) Schematic representation of the electrodes and channel of Layout 1, consisting of 1 channel of dimensions 60$\times$300 $\mu$m$^{2}$. (b) The same for Layout 2, consisting of 7 channels of the same width (20$\mu$m) and varying lengths (5, 10, 20, 50, 100, 200 and 500 $\mu$m). (c) A Schematic top view of the sample designed according to Layout 2. Except for the large Au gate electrode around the center of the sample, the Au contacts at the sides of the samples, and the channel area near the center, the whole surface is covered by a hard-baked photoresist layer (purple). For the pattern illustrated here, the contacts at the sides are connected to a Hall bar configuration for measuring the electronic properties of the channel area of SrTiO$_3$. A bias voltage $V_{\rm sd}$ between source and drain electrodes is applied, while simultaneously measuring the source-drain current $I_{\rm sd}$. The gate electrode (G), has an exposed area of more than $100$ times the area of the exposed SrTiO$_3$ channel. The potential difference $V_{\rm g}$ is applied between the gate and drain electrodes, while the gate current $I_{\rm g}$ is measured simultaneously.}
	\label{fig:Layouts}. 
\end{figure}

All of the sample processing was done on pristine undoped SrTiO$_3$ (001) crystal substrates with a miscut angle of $<0.1$ degrees, as obtained from Crystec GmbH. The contacts to the SrTiO$_3$ surface were defined by means of photolithography, as is shown in Fig.~\ref{fig:Micrograph}a. 
After development of the photoresist the patches on the surface onto which the electrodes were to be structured were first treated by Ar ion milling ($500$ V, $0.4$ Pa), such as to ensure low-resistance ohmic contacts between the metal electrodes and the SrTiO$_3$ surface. After this process, layers of $6$~nm Ti and $150$ nm Au were deposited by sputtering. In order to reduce leakage currents, in a second photolithography step the sample surface was covered by a separator layer (SL) in the form of an insulating photoresist  (purple in Fig.~\ref{fig:Micrograph}a), except for the electrical contact pads, the gate electrode, and the surface channel area. Chemical cleaning in an oxygen plasma ($10$ Pa, $13$ W) was used for removing any residuals of the photolithographic process on the bare SrTiO$_3$ surface. The resulting surfaces present flat terraces with step heights corresponding to the SrTiO$_3$ unit cell parameter of $0.391$~nm, as was verified by atomic force microscopy (AFM), Fig.~\ref{fig:Micrograph}b. We have tested patterns defined on the chip consisting of a single channel with multiple contacts, permitting four-probe measurements and voltage measurements transverse to the current direction (Layout 1, see Fig.~\ref{fig:Layouts}a), as well as patterns defining many channels of various lengths in two-probe configurations. Typically a channel is $20$~$\mu$m wide and between $10$ and $500$~$\mu$m in length (Layout 2, see Fig.~\ref{fig:Layouts}b).

After fabrication the sample was placed inside a glovebox under pure N$_2$ atmosphere  ($<0.1$~ppm O$_2$, H$_2$O), where it was heated to $120$~C for $1$ hour in order to remove adsorbates from the surface. The ionic liquid used here is N,N-diethyl-N-(2-methoxyethyl)-N-methylammonium bis(trifluoromethylsulphonyl)-imide (DEME-TFSI), which was kept stored in a closed bottle. Special care was taken in keeping the ionic liquid free from oxygen and water as much as possible. Before use, the bottle was opened in the glove box and the IL was pretreated by heating it to $60$~C for a period of $72$ hours. A metal needle was then inserted into the bottle in order to pick up a small droplet at the tip of the needle. This droplet was applied on top of the sample such as to cover the Au gate electrode and the channel area, as illustrated schematically in Fig.~\ref{fig:Layouts}c, followed by the transportation of the sample to the sample chamber of the Oxford Instruments Cryofree Teslatron cryogenic measurement system. Before starting the measurements this chamber was thoroughly flushed with high-purity He gas and pumped to $10^{-4}$~Pa to evacuate contaminations from the atmosphere before cool-down of the sample.

We have used two Keithley 2400 SourceMeters for the experiments, one of which was used for setting up a source-drain bias voltage of $V_{\rm sd}\le250$~mV while simultaneously measuring the source-drain current $I_{\rm sd}$. We verified that the results are independent of this setting using compliance values as low as $20$~mV. The initial, insulating state of SrTiO$_ 3$ has a resistance above our measurement limit of $25$~G$\Omega$. The other SourceMeter was used for applying the desired gate voltage $V_{\rm g}$, and for simultaneously monitoring the gate current $I_{\rm g}$. 

We have tested only positive gate voltages, leading to electron doping of the SrTiO$_3$ surface, in accordance with previous experiments. In order to minimize electrochemical carrier doping of the channel the gate voltages $V_{\rm g}$ were limited to a maximum of $3.2$~V \cite{Ueno2010}. Furthermore, the temperatures at which the gate potential was first applied, which we will refer to as the charging temperatures, were chosen to be close to the glass transition temperature $T_{\rm g} \approx 182$ K of DEME-TFSI \cite{Sato2004}.

\section{Experimental results}
\begin{figure}[!t]
	\includegraphics[width=0.6\textwidth]{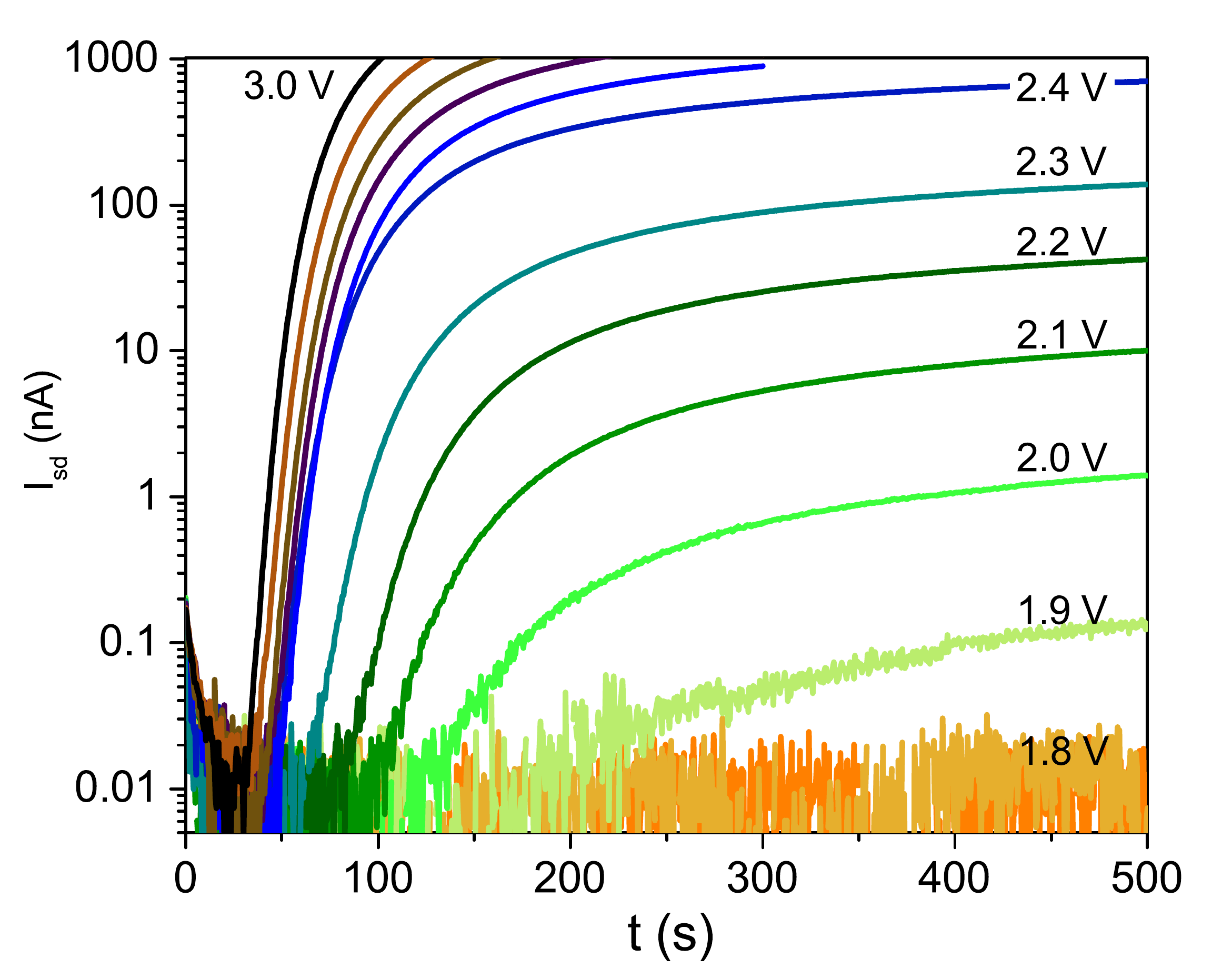}
	\caption{Time evolution of the source-drain current for a channel, 10$\times$20 $\mu$m$^{2}$, Layout 2, 
	at a bias voltage of $V_{\rm sd}=250$~mV. After stabilizing the temperature at $T=195$~K and at $V_{\rm g} =0$ 
	the gate voltage $V_{\rm g}$ is switched at $t=0$ instantaneously to the values indicated by the labels for each of the curves. 
	A time $t_{\rm d}$ elapses before the current suddenly rises above the noise floor of about $10$~pA, and the current increases nearly 
	exponentially. 
	 }\label{fig.Isd}
\end{figure}

We start the experiments by stabilizing the sample at a temperature $T$ above the glass temperature of the IL, after which at $t=0$ the gate potential $V_{\rm g}$ is switched from zero to a chosen fixed value, and the gate current is recorded simultaneously with the channel conductance. We observe the following: After switching the gate potential a waiting time $t_{\rm d}$ elapses before we observe any measurable response in the conductance of SrTiO$_3$, as shown in Fig.~\ref{fig.Isd} for $T=195$~K and several settings of $V_g$. \footnote{The small current signal decaying immediately after $t=0$ is part of the gate current that flows through the source contact. The gate current is discussed in more detail further down in the text.} When the gate is switched back to zero the decay of the conductance  takes place on a much shorter time scale. We introduce a waiting time of at least $10$~minutes before a new gate voltage is set.

\begin{figure}[!t]
	\includegraphics[width=0.6\textwidth]{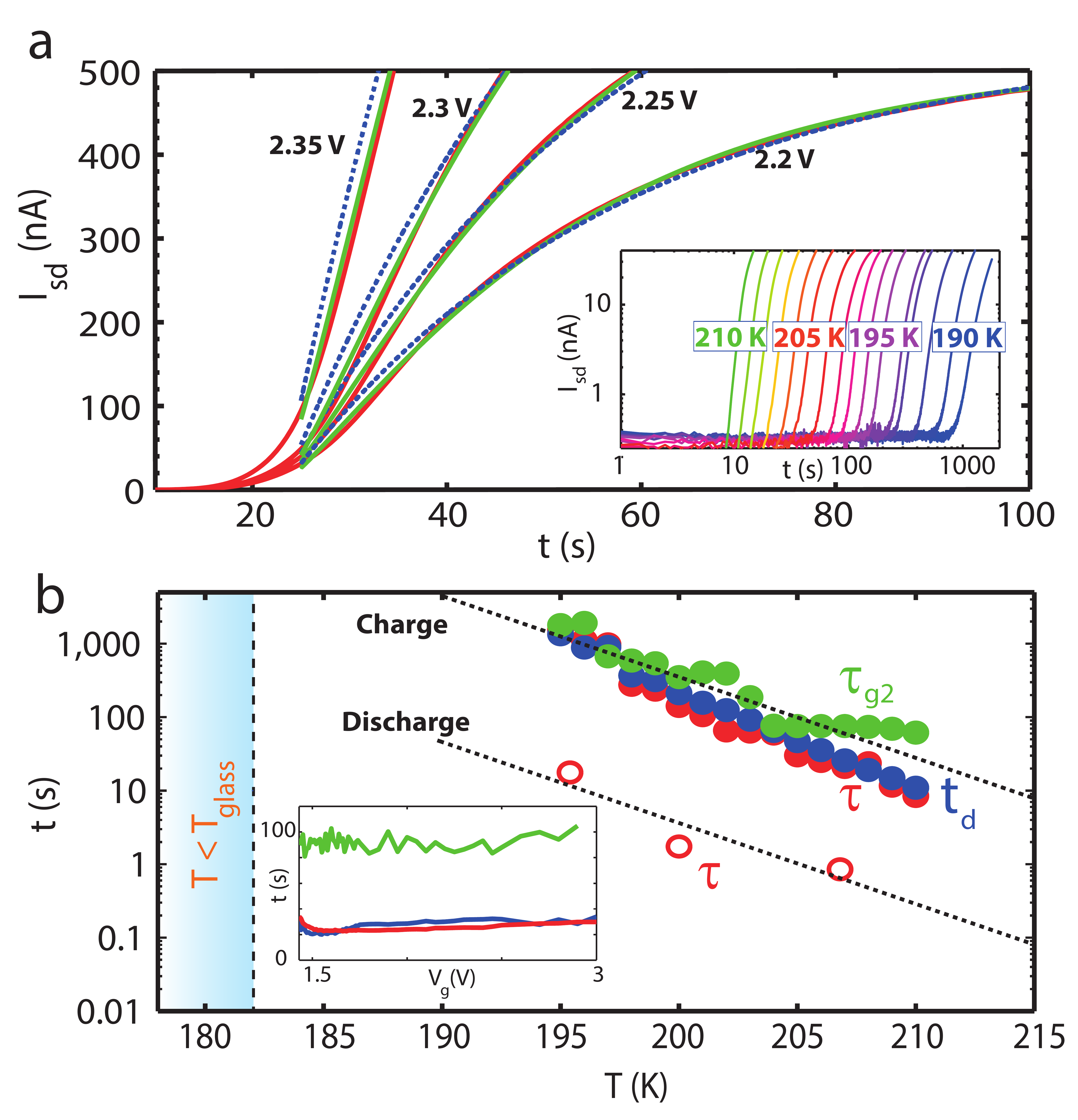}  
	\caption{Time delay in the response of the conductance of a SrTiO$_3$ channel (10$\times$20 $\mu$m$^{2}$, Layout 2) upon application of a gate potential. 
	(a) Source-drain current as a function of time after switching to a fixed gate voltage as indicated, for a charging temperatures of $205$~K. 
	The blue dotted curves show the single-exponential fits (see text). The green curves give the fits based on the percolation model. The inset shows the source-drain current on a log scale for a range of temperatures
	and for $V_{\rm g}=2.5$~V. 
	(b) Time delay $t_{\rm d}$ (blue) and time constants $\tau$ (red) as obtained from the fits to the time-evolution of the source-drain current in (a) for $V_{\rm g}=2.5$~V, 
	plotted as a function of the charging temperature. 
	For a few temperatures we also show the discharge time constants, which are much faster. The green points represent the time constant $\tau_{\rm g2}$ for the decay of the gate current. 
	The time constants increase nearly exponentially when approaching the glass temperature of the ionic liquid at $182$~K, as illustrated by the straight dashed lines. 
	The inset shows that the time constants are nearly independent of the gate voltage, for $T=205$~K, 
	where we use the same color coding as in the main panel.}
	\label{fig.time-delay}
\end{figure}

The delay time of the main signal strongly increases for lower temperatures (as illustrated for a different sample in Fig.~\ref{fig.time-delay}, Layout 2) and becomes very long ($>1000$~s) when the temperature approaches the glass transition temperature of about $182$~K of the ionic liquid. 
After the conductivity onset, the current approaches a nearly exponential increase, as shown by the blue dotted curves in Fig.~\ref{fig.time-delay}a. The current $I_{sd}(t)$, here plotted on a linear scale, is closely described by a single exponential function $I_{sd}(t) = I_0 (1 - \exp(-(t-t_{\rm d})/\tau)$, which defines our time constants $t_{\rm d}$ and $\tau$. These time constants are plotted as a function of temperature in Fig.~\ref{fig.time-delay}b, showing that the times rapidly grow for temperatures approaching the glass transition, and that the two times are of similar magnitude. 

\begin{figure}[!t]
	\includegraphics[width=0.7\textwidth]{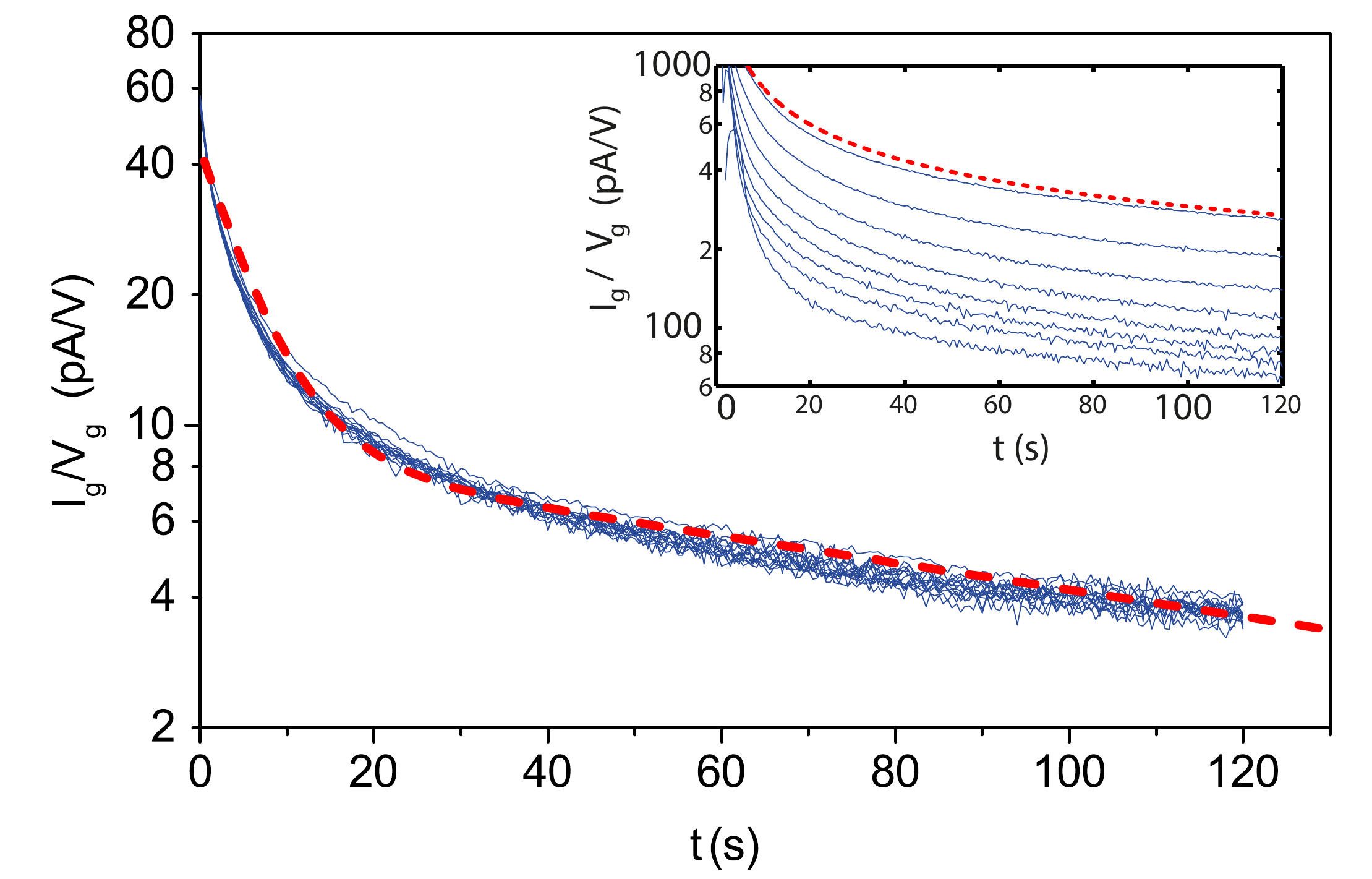}
	\caption{Time evolution of the gate current $I_{\rm g}$, scaled by the gate potential, upon application of 15 different gate potentials 
	ranging from
	$1.4$~V to $3.0$~V at a charging temperature of $205$~K. The curves perfectly collapse to a single smooth time evolution, which is closely
	described by a combination of two exponential functions with time constants $ \tau_{\rm g1}$ and $\tau_{\rm g2}$ (red broken curve). The inset shows the curves between $2.3$~V to $3.0$~V at a charging temperature of $220$~K, typical of experiments performed at a minimum pressure of $1$ Pa. In such circumstances, the curves no longer are described by a universal time dependence, and show clear $V_{\rm g}$ dependence. The individual curves are better described with a Faradeic term. The red, broken curve is a single $t^{-1/2}$ term fit of the data belonging to $V_{\rm g}=3.0$ V ($R^{2}=0.9814$).}
	\label{fig.gate-current}
\end{figure}

The gate current, on the other hand, switches instantaneously, followed by a smooth decay, as shown in Fig.~\ref{fig.gate-current}. We have scaled each of the gate currents by the applied gate potentials, ranging from $1.4$~V to $3.0$~V, and the resulting curves perfectly collapse to a universal time dependence. The lack of voltage dependence in the scaled gate current provides strong support for a purely electrostatic charging process. The curves in Fig.~\ref{fig.gate-current} are closely described by a combination of two exponentially decreasing functions, with time constants $\tau_{\rm g1}$ and $\tau_{\rm g2}$, as illustrated by the red broken curve. The absence of long-time tails in the gate current with $\sqrt{t}$ dependence further testifies for the absence of electrochemical contributions to the charging process \cite{Ueno2010}. We find that the precautions taken to prevent oxygen and water from interacting with the ionic liquid are critical for this result. For example, when pumping the sample chamber of the cryogenic system to only $1$~Pa we observe strong deviations from purely exponential decay of the gate current, the scaling of the gate current with gate voltage breaks down (Fig.~\ref{fig.gate-current} inset), and the saturation value of the source-drain current becomes only weakly dependent on the gate potential.

The time constant $\tau_{\rm g2}$ is also shown in Fig.~\ref{fig.time-delay}b, and has a similar magnitude and temperature dependence as found for $t_{\rm d}$ and $\tau$. The time constants are nearly independent of the gate potential to within 20\%, as shown in the inset. 

\section{Discussion}
The fact that all time constants, $t_{\rm d}$, $\tau$, and $\tau_{\rm g2}$, 
have very similar values and that they depend in the same way on temperature suggests that the processes are intimately linked, and are dominated by the ionic conduction in the liquid. The ionic conductivity becomes very small close to the glass transition and limits the charging time of the total capacitance $C_{\rm tot}$. The ionic conductivity can be described by the Vogel-Fulcher-Tamman (VFT) equation \cite{Ohno2005},
\begin{equation}
\rho=A\sqrt{T}\exp{\frac{B}{T-T_{0}}}. \label{eq.VFT}
\end{equation}
Here, $A$ and $B$ are constants related to the ion density and the activation energy for ion transport, respectively, and $T_{0}$ is the ideal glass transition temperature. The temperature dependence of the time constants in Fig.~\ref{fig.time-delay}b agrees with this nearly exponential dependence, as indicated by the broken lines in the figure, although the temperature interval explored here is not wide enough for testing the full expression.

The time delay in the source-drain current $I_{\rm sd}$ and the fact that the gate current in Fig.~\ref{fig.gate-current} does not show a single-exponential decay as expected for the simple charging of a capacitor could be due to the charging properties of the IL. Yuan {\it et al.} \cite{Yuan2010} have analyzed the dynamic capacitance of the system of the DEME-TFSI ionic liquid in contact with a ZnO surface as a function of temperature and frequency. They find that the system can be represented by two resistor-capacitor systems in series. One of these represents the charging of the surface capacitance, and the other the geometric response of the ionic liquid. However, the latter produces a response on time scales that are two or three orders of magnitude shorter than those observed for the initial fast decay of the gate current, as can be read from Fig. 5 in Ref. \cite{Yuan2010} and such processes, if present, would not be resolved on the timescale of our experiments. In our case the two time scales find a natural explanation in terms of the voltage dependence of the surface capacitance of SrTiO$_3$.

Given the fact that SrTiO$_3$  is a wide band gap insulator the delay time in the source-drain current $I_{\rm sd}$ can be attributed to the role of the threshold potential $V_{\rm th}$ required for bringing the electrochemical potential at the surface of SrTiO$_3$ to the edge of the conduction band. After switching on the gate voltage the potential at the SrTiO$_3$ surface is expected to rise exponentially in time. At equilibrium the potential in the ionic liquid near the surface of SrTiO$_3$ is given by the division of the potential over the series connection of the capacitance at the gate and the capacitance to the outside world. Since the former is orders of magnitude larger, the potential at the surface reaches nearly the full gate potential. In the process of building up this surface potential as a function of time the conduction band of  SrTiO$_3$ is locally pulled down until, at a threshold potential $V_{\rm th}$ the chemical potential of the electrons in the Au leads is aligned with the bottom of the conduction band. This condition is met after a time delay  $t_{\rm d}$, and from this moment on a two-dimensional electron system (2DES) can develop at the surface of SrTiO$_3$. As the ionic surface charging continues towards saturation the density of carriers in the 2DES grows and the conductivity increases.
This interpretation of the delay time is consistent with the fact that we observe no substantial conductivity for gate voltages smaller than $1.8$~V.

The integrated gate current after saturation gives the total charge $Q$ and  we find that this scales linearly with the gate potential, in agreement with the scaling in Fig.~\ref{fig.gate-current}. The proportionality constant between $Q$ and $V_g$ gives the total capacitance, $C_{\rm tot}=Q/V_{\rm g}$, for which we obtain $1.4$~nF, for the sample configuration of the data in Figs.~\ref{fig.time-delay} and \ref{fig.gate-current}. This capacitance is  larger than expected for a capacitance associated with an individual channel. For the channel with size $20\times10 \mu$m$^2$ used for the measurements presented above, and a specific capacitance at high gate voltage at the SrTiO$_3$ surface of $13$~$\mu$F/cm$^2$ \cite{Tsuchiya2015}, the expected capacitance is $26$~pF. The larger value of the capacitance obtained from the gate current reflects the fact that the capacitance associated with the gate current is the total capacitance of the ionic liquid droplet to the outside world and includes the capacitive coupling to the other channels on the sample. Estimates based on the total exposed area of SrTiO$_3$ gives $2.3$~nF, which agrees within a factor of two with observation. 

The fact that the capacitance is dominated by the interface with SrTiO$_3$ implies that the capacitance must be voltage dependent. The delay time observed in the source-drain current implies that this capacitance switches to a larger value as soon as the potential at the surface brings the Fermi energy up to the conduction band edge. We take this effect into account by a simplified model through a combination of two time constants, $I_g(t) = I_1 \exp(-t/\tau_{\rm g1}) + I_2 \exp(-t/\tau_{\rm g2})$, and the fit to this expression is shown by the broken curve in Fig.~\ref{fig.gate-current}. The second time constant is the one associated with the charging of the surface capacitance in the regime of a finite conducting surface charge density. The cross over between the two regimes occurs at a time that is similar to the delay time observed in the source-drain current. The initial time constant $\tau_{\rm g1}$ is attributed to the residual capacitance, dominated by the Au contacts to the channels.

The discharge curve for the source-drain current, recorded when switching the gate potential back to zero, is again closely described by a single exponential. This represents the relaxation of the EDL and we observe that this process is much faster than the build-up of the double layer, as shown in Fig.~\ref{fig.time-delay}b. As was discussed in Ref.~\cite{Li2017} the difference in charging and discharging times can be understood as the competition between two driving forces. When charging the EDL the cations are driven towards the interface by the electric field, but the concentration gradient drives them in the opposite direction. When discharging, the electric field of the charge layer works in the same direction as the concentration gradient, resulting in a larger combined driving force for the current. From these observations we conclude that the ionic conductivity obtained from the process of charging and discharging of the EDL may deviate from the bulk conductivity.

\begin{figure}[t!]
	\centering
	\includegraphics[width=0.6\textwidth]{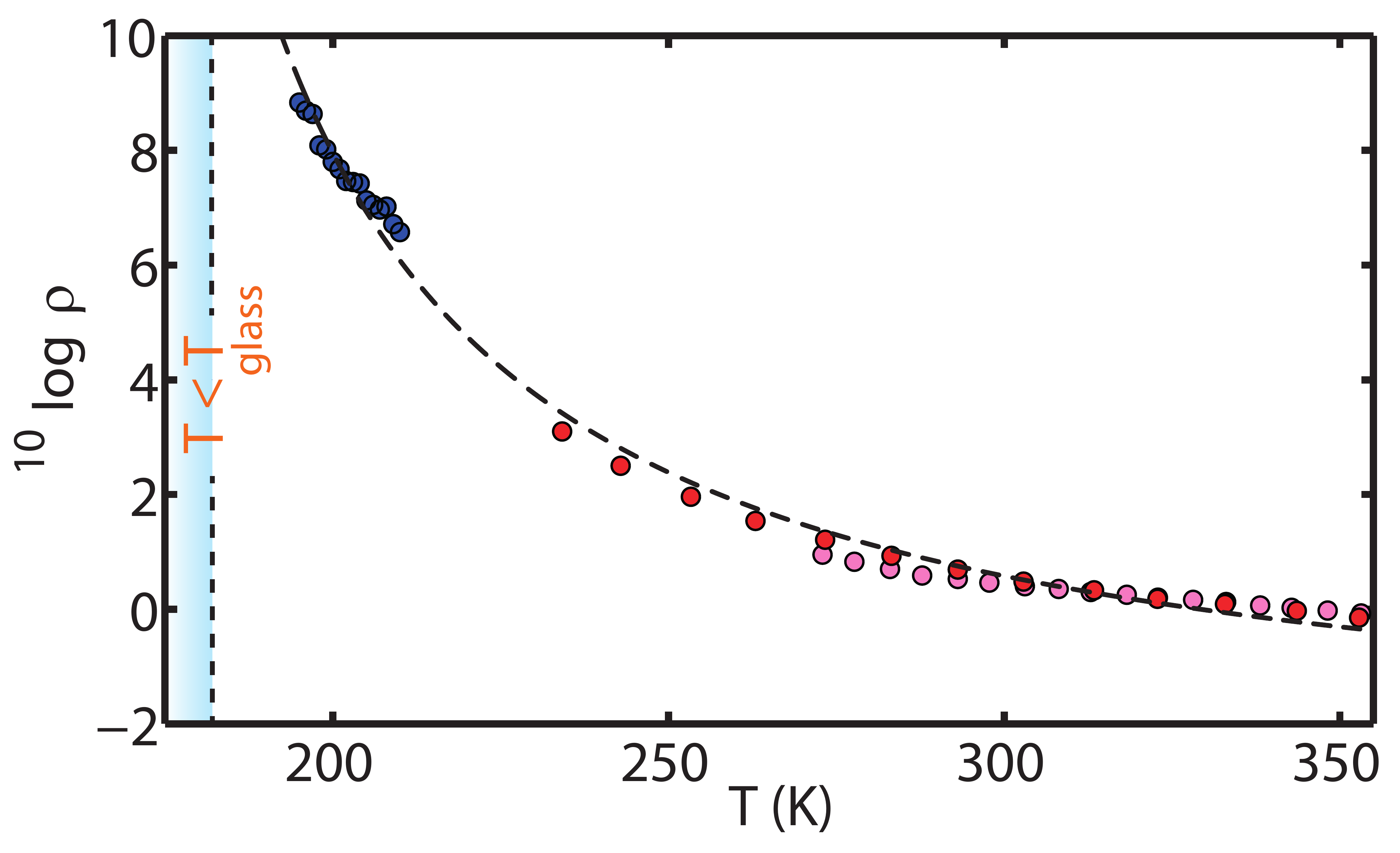}
	\caption{Plot of the ionic liquid resistivity as a function of temperature. The blue points are obtained in this work as calculated from the
	observed delay times. Red and pink points represent literature values for $\rho$, as taken from Refs. \citenum{Sato2004} and
	\citenum{Hayamizu2010}. The fit curve is based on the Vogel-Fulcher-Tamman equation. The fit parameters are: $T_{0}= 151\pm7$~K,
	$A=4.8\times 10^{-3}$ Sm$^{-1}$K$^{-0.5}$ and $B = 1.3\times 10^{3}$ K ($R^{2}= 0.994$).}
	\label{fig.rho}
\end{figure}

Nevertheless, we can use the experimental values for $\tau$ for estimating the ionic resistivity $\rho$ taking into account the geometry and size of the electrodes according to Layout 2. We take the channels as approximately one-dimensional, and combine the contribution to the charging current from all channels on the sample. The resistivity can then be approximated by $\rho=\pi R L/\ln(W/a)$, where $R=\tau/C$ is the resistance obtained from the time constant and the capacitance, $L=910\, \mu$m is the total length of the channels, $a=10\, \mu$m is half the width of the channels, and $W\simeq 1$~mm is the radius of the ionic liquid droplet. The values for $\rho$ thus obtained are accurate to within a factor of two as limited by our knowledge of the geometry of the droplet (represented by the errorbars), and are plotted in Fig.~\ref{fig.rho} along with the known literature values \cite{Sato2004,Hayamizu2010}. The curve fit is based on the VFT equation (\ref{eq.VFT}) \cite{Ohno2005}. The fit parameter $T_{0}$ is the ideal glass transition temperature, which is typically $30$ to $50$~K lower than the $T_{\rm g}$ measured by means of differential scanning calorimetry \cite{Ohno2005}, as is the case for DEME-TFSI \cite{Sato2004}. In our case the VFT function gives a good fit, with $T_{0}=151\pm$7~K, demonstrating that our interpretation of the delay time produces values for the ionic liquid conductivity that are in good agreement with the available literature data.
Note that our data points appear to be slightly above an extrapolation of the literature values, which may result from the fact that our resistivity data are obtained from the properties of the EDL and thus be influenced by the concentration gradient.

In order to connect the charging/discharging dynamics in the ionic liquid with its effect on the electrical conductivity of the channel we now turn to the properties of the 2DES, notably its conductivity extrapolated to long times, $\sigma_\infty$.  Despite the metallic characteristic at high charging voltages \cite{Ueno2008}, the 2DES cannot be viewed as a normal metal because we find that the conductivity strongly depends on the length of the channels, decreasing for channel lengths running from  $10$ to $500$~$\mu$m. As another indication of this anomalous character the conductivity of the 2DES remains well below the conductance quantum, $\sigma_\infty \ll e^2/h$. These observations are consistent with an interpretation of the conductivity in terms of Anderson localization and the formation of percolation networks of conducting paths \cite{Li2012}. 

This observation appears to be contradicting the nearly perfect exponential growth of conductance with time seen in Fig.~\ref{fig.time-delay}. In a 2DES percolation model the conductance is expected to be controlled by the charge density $n$ according to \cite{Adam2008}, 
\begin{equation}
\sigma = A \left( n- n_{\rm c} \right)^{4/3}, \label{eq.sigma}
\end{equation}
where $A$ is a system specific constant and $n_{\rm c}$ is the two-dimensional critical density for forming a percolation path.
Assuming a sharp conduction band edge, the charge density at the surface is controlled by the local electrostatic potential $V$ created by the ionic liquid at the surface of the SrTiO$_3$ crystal, as
\begin{equation}
n(V) = c \left( V - V_{\rm th} \right),   \label{eq.n}
\end{equation}
where $V_{\rm th}$ is the threshold voltage determined by the position of the conduction band edge, and $c$ is the capacitance per unit area. From the discussion above we conclude that after the delay time the local potential $V$ follows a simple exponential law for the charging of a capacitor $V(t) = V_{\rm g}(1-\exp(-t/\tau))$. Combining this with Eqs.~(\ref{eq.sigma}),(\ref{eq.n}) we expect a time evolution for the conductivity,
\begin{equation}
\sigma(t) = \sigma_\infty \left( \alpha -\exp(-t/\tau)) \right)^{4/3}. 
\end{equation}
Here, $\sigma_\infty = A ( c V_{\rm g})^{4/3}$, and $\alpha=1-V_{\rm th}/V_{\rm g} - n_{\rm c}/(cV_{\rm g})$ is a constant of order unity. This functional form fits the observed time dependence very well, as demonstrated by the green solid curves in Fig.~\ref{fig.time-delay}.

The close match between the fit and the observed time dependence of the conductance supports the interpretation that the conductance is just controlled by the local electrostatic potential of the ionic liquids at the surface. 

\begin{figure}[t!]
	\centering
	\includegraphics[width=0.6\textwidth]{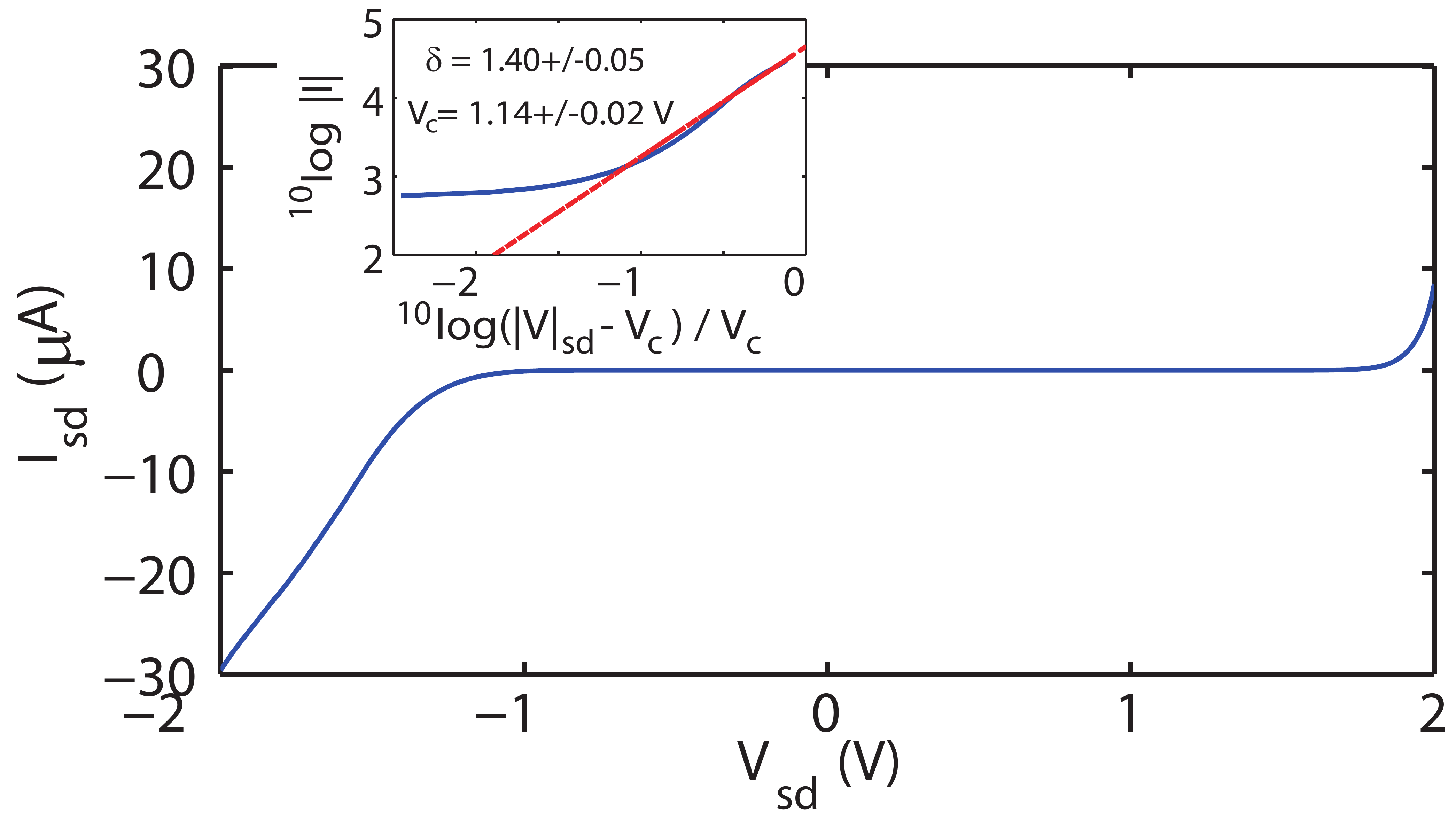}
	\caption{I(V) of the channel for $V_{\rm g}=2.5$ V taken at $T=20$K. The curve is non-linear and can be described by the RRTN percolative model, where $I_{\rm sd}\propto(V_{\rm sd}-V_{\rm c})^{\delta}$. The inset shows the corresponding log-log plot of the negative $V_{\rm sd}$ part of the data. The red, broken line shows the fit to the data with $V_{c}=1.2$ V, and $\delta=1.30\pm0.02$.
	}
	\label{fig.IV}
\end{figure}

For 2D percolative systems such as random-resistor-tunneling-networks \cite{Sen2009}, certain I(V) characteristics are expected. When a voltage is applied between the source and drain electrodes, no current flows below a critical voltage $V_{c}$. For $V>V_{c}$ a certain scaling behavior is to be expected, according to $I_{sd}\propto(V-V_{c})^{\delta}$ \cite{Rimberg1995}, for which the value $\delta$ corresponds to the same $4/3$ \cite{Sen2009}, also found in SrTiO$_{3}$ \cite{Mingyang2012}. We find similar, non-linear behavior at low temperatures for gated channels (Fig. ~\ref{fig.IV}). The channel gated at $V_{\rm g}=2.5$ V shows scaling behavior with $\delta=1.30\pm0.02$ and $V_{\rm c}=1.2$ V at $T = 20$K.

Any electrochemical influence of the surface charge density would involve diffusion of the reaction species towards and away from the surface and is expected to show up as a time dependence $t^{-1/2}$ \cite{Zhou2012,Ueno2010}. This would be seen as a lack of saturation at long times, which we encounter in our experiments when we relax the conditions for treatment of the ionic liquid, so that oxygen (or water) contamination starts playing a role. 
For example, when evacuating the sample space at the start of the experiments to only $1$~Pa we observe a break down of the scaling in Fig.~\ref{fig.gate-current} and the 2DES conductivity does not saturate. Remarkably, poorer vacuum conditions often lead to higher maximum 2DES conductivities, but for longer gating times the conductivity starts degrading and becomes dependent on the gating history.

Apart from the methods employed in this work to illustrate the difference between electrostatics and electrochemicy, other methods can also be employed to further illustrate this contrast through enhancement of electrostatic gating. An example of this is the usage of monolayer separator layers such as hexagonal boron nitride \cite{Gallagher2015} or other insulating materials to prevent direct contact between the ionic liquid and the surface of \ce{SrTiO3}. Although such although the increased double layer distance limits the polarization and changes the Coulomb scattering and mobility properties of the 2DES, measurements of the onset of conduction with such a separator layer in place could provide a further means of minimizing any possible contribution of electrochemistry to the gating process here.  

In the analysis of the time dependence we have ignored delays in charging times due to the finite conductivity of the 2DES. In analyzing time delays similar to those reported here for experiments on SrTiO$_3$ with a solid electrolyte Tsuchiya {\it et al.} \cite{Tsuchiya2015} proposed a model of charge build up that evolves gradually over the surface of the solid, initiated from the contact electrodes. Initially the SrTiO$_3$ surface is assumed to be a perfect insulator. Near the electrodes the build up of an electrostatic potential in the ionic liquid pushes the chemical potential to the conduction band edge, allowing local charging of the SrTiO$_3$ surface. This charged surface then serves as the contact electrode for the next section of the surface. 

If this model were correct the time delay would depend linearly on the channel length. We have tested this in two ways. First, for fixed channel length in Hall bar configuration we measured the conductivity across the channel at two points, using small capacitors to connect to the side contacts and an ac lock-in resistance measurement technique. The coupling by small capacitors prevents the side contacts from forming nucleation points for the formation of a 2DES. When switching the dc source-drain bias we find that the conductance along the channel and at two points across the channel develop nearly simultaneously. The noise on the ac signal limits our accuracy in determining the delay times,  but they are equal to within $20$\%. 
In addition, we tested on another sample a series of channels in $2$-point configuration with lengths of $10$, $20$, $50$, $100$, $200$ and $500$~$\mu$m. The observed time delays show a variation by about $50$\%, but no systematic systematic dependence. In fact, the shortest channel showed the longest delay time.  We conclude that the charging of the surface occurs nearly homogeneously, and we tentatively attribute the small variation in delay times to variation in the geometries and positions on the sample.

The absence of a length dependence in the delay times suggests that the conductivity of the 2DES is not a limiting factor in the formation of the EDL. We picture the charging process as follows. When a gate potential is applied a standard EDL forms at the surfaces of the gate and at the Au contact electrodes. Since the area of the gate is much larger, the voltage drop concentrates at the surface of the small source and drain contacts. Towards equilibrium the potential would assume a homogeneous level inside the ionic liquid, and this would approach the potential of the gate. Since the surface of the SrTiO$_3$ crystal is in contact with this ionic liquid it feels the rising of the potential of the ionic liquid towards that of the gate over its entire surface. Only when this potential brings the conduction band edge to the level of the Fermi energy the  EDL starts to form, and electrons flow in from the electrodes. At that moment the conductance is already well above the conductance of bulk SrTiO$_3$ in equilibrium. When the resistance of the 2DES drops below our detection limit of about $25$ G$\Omega$, the estimated RC time for a channel of $10\times20\,\mu$m, using the quoted specific capacitance of $13$	~$\mu{\rm F/cm}^2$, is only $0.6$~s, much smaller than the observed time delays. We conclude that the observed time dependence is entirely dominated by the ionic conductivity of the IL.

In conclusion, we have demonstrated that IL-gating on insulators, {\it i.c.} SrTiO$_3$, at charging temperatures close to the glass temperature $T_{\rm g}$ produces a delayed transition to a conducting surface state. The delay time is strongly dependent on the temperature and diverges near $T_{\rm g}$. The time delay and time evolution of the conductance can be described by a process of homogeneous charging of the surface, and is determined by the band gap of the insulator. The conductance that results is consistent with a percolation model of transport. We provide a method of distinguishing between electrochemical and electrostatic processes by using the scaling behavior of the gate current, which is of crucial importance in this rapidly developing field of research. We find no evidence of charge doping by electrochemical reactions when charging close to the glass temperature of the IL, provided we maintain strict conditions during handling of the ionic liquid.

\begin{acknowledgement}
This work is part of the research program of the Foundation for Fundamental Research on Matter (FOM), which is financially supported by the Netherlands Organisation for Scientific Research (NWO). The authors gratefully acknowledge generous support in the experiments and analysis by Jan Aarts, Stefano Voltan and Serge Lemay.
\end{acknowledgement}

\bibliography{STO_AdP}
\bibliographystyle{achemso}

\end{document}